\documentclass{aastex7}

\begin{document}

\title{Moving Plasma Structures and Possible Driving Mechanisms of Solar Microflares Observed with High-Resolution Coronal Imaging}

\author{Qingmei Wang}
\affiliation{Key Laboratory of Colleges and Universities in Yunnan Province for High-energy Astrophysics\\
Department of Physics, Yunnan Normal University\\
 Kunming 650500  \\
 China}
\affiliation{Yunnan Observatories, Chinese Academy of Sciences\\
396 Yangfangwang, Guandu District \\
 Kunming, 650216\\
 China}
 \email{}

\author[0000-0002-5302-3404]{Yi Bi}
\affiliation{Yunnan Observatories, Chinese Academy of Sciences\\
396 Yangfangwang, Guandu District \\
 Kunming, 650216\\
 China}
 \affiliation{Yunnan Key Laboratory of Solar Physics and Space Science\\
 396 Yangfangwang, Guandu District \\
 Kunming ,650216 \\
 PR China \\ }
\email[show]{biyi@ynao.ac.cn}

 \author{Hongfei Liang}
\affiliation{Key Laboratory of Colleges and Universities in Yunnan Province for High-energy Astrophysics\\
Department of Physics, Yunnan Normal University\\
 Kunming 650500  \\
 China}
\email[show]{3632@ynnu.edu.cn}

\author[0000-0003-3462-4340]{JiaYan Yang}
\affiliation{Yunnan Observatories, Chinese Academy of Sciences\\
396 Yangfangwang, Guandu District \\
 Kunming, 650216\\
 China}
 \email{}
  
 \author{Liufan Gong}
\affiliation{Yunnan Observatories, Chinese Academy of Sciences\\
396 Yangfangwang, Guandu District \\
 Kunming, 650216\\
 China}
 \email{}


\begin{abstract}

Solar microflares are ubiquitous in the solar corona, yet their driving mechanisms remain a subject of ongoing debate. Using high-resolution coronal observations from the Solar Orbiter's Extreme Ultraviolet Imager (EUI), we identified about a dozen distinct moving plasma structures (hereafter, `` tiny ejections'') originating from the centers of three homologous microflares out of four successive events. These tiny ejections propagate roughly perpendicular to the flaring loops. They often originate as dot-like structures with a length scale of approximately $10^{3}$ km. While these initial dot-like shapes are observable in EUI images, they remain undetectable in the images captured by the Atmospheric Imaging Assembly onboard the Solar Dynamics Observatory. As they propagate, these dot-like structures consistently evolve into loop-like formations, possibly due to the heating of the surrounding magnetic field. Rather than being generated by a series of flux rope eruptions, the tiny ejections appear to result from small-angle magnetic reconnections within a bipolar field. Thus, the microflares associated with these ejections may be driven by magnetic reconnection within braided fields, a process similar to the proposed nanoflare mechanism and distinct from the standard large-scale flare model.

\end{abstract}

\keywords{Solar flares (1496), Solar magnetic reconnection (1504), Solar coronal (1483) }


\section{Introduction}

Microflares are localised, small-scale energy release phenomenon on the Sun induced by magnetic reconnection (e.g.,\citealt{2004ApJ...612..530Q,2008ApJ...686..674N,2010ApJ...712L.111J,2017ApJ...845..122G}), usually manifesting as brightening of extreme ultraviolet (EUV)(e.g.,\citealt{Chen_1999,2009ApJ...692..492B,2010ApJ...724..640C}), soft X-ray (SXR)(e.g.,\citealt{2000ApJ...534..482T,Shimizu_2002,2010ApJ...720.1136K}) and hard X-ray (HXR)(e.g.,\citealt{2004ApJ...612..530Q,2008ApJ...686..674N,2009ApJ...692..492B}) radiation enhancement. In the GOES classification, they are usually classified as A-Class and B-Class events, with some events exceeding the GOES sensitivity limits. They originate as transient, localised thermal plasmas in the chromospheric environment at an altitude of approximately 2--10 Mm above the solar photosphere, i.e., they typically occur in the chromosphere and the lower corona and persist for a duration of 3--10 minutes, releasing magnetic energy within a range of $10^{26}$--$10^{29} $ erg \citep{2022ApJ...930L...7L}. 

It is well known that the energy distribution from microflares to major flares obeys a power-law \citep{2002ApJ...572.1048A}. Although the energies released during the eruption of microflares are small, they are much more numerous compared to the major flares.  This may indicate that the total energy deposited into the corona by these small-scale flares is even higher than that of the large-scale flares, and thus is also suspected to be a possible cause of coronal heating. However, this conclusion remains controversial because whether microflares contribute to coronal heating depends on whether their power-law slope is greater or less than 2. If the power-law slope is above 2, small-scale flares could indeed contribute significantly to coronal heating \citep{1991SoPh..133..357H,2022ApJ...934L...3A}.

The consistency of physical processes and energy properties between microflares and larger flares remains an open question. While \cite{Inglis_2014} suggested that microflare thermal energy might originate from mechanisms other than the standard chromospheric evaporation model (e.g., direct heating, accelerated protons, plasma waves, or direct current fields), several observations point towards a more unified picture. The presence of the Neupert effect \citep{1991BAAS...23R1064H}, where the time derivative of soft X-ray (SXR) emission mirrors the hard X-ray (HXR) light curve in microflares \citep{2022A&A...659A..52S}, coupled with a similarity in the ratio of non-thermal to thermal energy between microflares and large flares \citep{2020ApJ...891L..34G}, strongly suggests a significant role for chromospheric evaporation in microflares.  This accumulating evidence points to consistency between microflare and large flare energy release mechanisms, warranting further investigation into the underlying physical processes.

A large number of studies of microflares have shown that microflares are multi-scaled self-similarity to major flares \citep{1996ApJ...460.1034F,2008ApJ...677.1385C,2011SSRv..159..263H}. \cite{2015PASJ...67...40J} suggests that the reconnection between the newly emerged magnetic field and the pre-existing magnetic field leads to microflares. Recently, \cite{2022ApJ...930L...7L} found that microflares can also result from tether cutting (as shown in \citealp{2015NatCo...6.7598S}) or fan-spine reconnection. 

Microflares are often accompanied by jets \citep{1992PASJ...44L.173S,2008A&A...491..279C,1996PASJ...48..123S,2021RSPSA.47700217S,2023A&A...670A..56B}. The term ``jet" typically refers to collimated, bundle-like jet plasma streams that follow straight or slightly oblique magnetic field lines \citep{2016SSRv..201....1R}. \cite{Moore_2010} distinguished between two categories of jets: standard jets and blowout jets based on their different physical properties. Standard jets conform to the emerging flux interchange reconnection model \citep{1977ApJ...216..123H,1989ApJ...345..584S,1992PASJ...44L.173S,1996PASJ...48..353Y}. In contrast, blowout jets possess sufficient free energy to drive the jet eruption due to the high degree of shear and twist in the magnetic field of the arch core.

The jets that are perpendicular to the coronal structures were reported by \cite{2017ApJ...841L..13C}, who found that the so-called subjets, accompanied by solar tornadoes, are much smaller than typical jets and may result from magnetic reconnection that occur in coronal loops \citep{1988ApJ...330..474P,2008A&A...485..837B}, which differ from the physical processes associated with usual jets. \cite{2021NatAs...5...54A} reported smaller-scale jet structures that are transverse to the coronal loops, which they termed nanojets due to the accompanying energy release being at the level of nanoflares. Nanojets may be ubiquitous \citep{2021NatAs...5...54A,2021A&A...656A.141P,2022ApJ...934..190S,2022ApJ...938..122P} and are considered to result from magnetic reconnection among the braiding field within the coronal loops, significantly contributing to coronal heating. However, few studies have hypothesized or provided evidence that such a driving mechanism triggers larger energy-level eruption activities.

Recently, by the utilisation of high spatial and high cadence resolution time series observations from the Extreme Ultraviolet Imager (EUI; \citealt{2020A&A...642A...8R}) on board the Solar Orbiter (SO; \citealt{2020A&A...642A...1M}) has facilitated significant advancements in the field of solar physics, particularly in the study of small-scale transient phenomena. One such phenomenon, observed in the quiet corona by the EUI, is the small-scale transient brightening driven by magnetic reconnection, which has been termed a 'campfire' \citep{2021A&A...656L...4B}.
Furthermore, \cite{2021ApJ...918L..20H} defined a specific subset of campfires, known as coronal microjets. The triggering mechanism of these coronal microjets is consistent with the characteristics observed in classical jets during interchange reconnection events \citep{2016SoPh..291.1357W}. Observations from the EUI also reveal that the coronal magnetic field undergoes complex braiding and untangling activities, manifesting as nanoflares that release heat into the corona \citep{2022A&A...667A.166C}.

In this letter, we used observations from SO/EUI near its fourth perihelion to investigate microflares accompanied by a series of moving plasma structures that are perpendicular to the flaring loops. We refer to these moving  plasma structures associated with microflare as tiny ejections. These findings support the idea that magnetic reconnection within the braiding field could contribute to energy release at the microflare level. We provide  an overview of the data in Section \ref{sec:analysis}. Section \ref{sec:Event} presents the observations of the events. In Section \ref{sec:Conclusions}, we discuss our findings and draw conclusions.

\section{Observation and data analysis} \label{sec:analysis}
We utilized 174 \AA\ imaging data of the quiet region of the Sun near the center of the disk, taken by $EUI/HRI_{EUV}$ on the Solar Orbiter (SO) between 00:38 and 00:58 UT on 2022 March 8. The specific location of the event is at latitude-longitude coordinates ($15.7^{\circ}$, $-14.9^{\circ}$), with the SO positioned approximately 0.48 au from the Sun. The $EUI/HRI_{EUV}$ plate scale is 0.492{\arcsec}, with a temporal cadence of 5 seconds \citep{2023A&A...675A.110B}. During these observations, one pixel corresponded to a distance of about 172 kilometers on the Sun. Additionally, we employed data from the Atmospheric Imaging Assembly (AIA; \citealt{2012SoPh..275...17L}) on board the Solar Dynamics Observatory (SDO; \citealt{2012SoPh..275....3P}) to obtain full-disk images with a temporal cadence of 12 seconds and a pixel size of 0.6{\arcsec}. In this case, one pixel corresponds to a projection distance of approximately 400 km on the solar disk. The separation between SO and SDO in heliographic longitude is about $30^{\circ}$. In order to avoid confusion in subsequent analysis, the time of the EUI images was adjusted to that at 1  au, due to the discrepancy in the heliocentric distance of SO and SDO.

 \section{Observations of microflares in coronal magnetic field structures} \label{sec:Event}
\subsection{Tiny ejections link to microflares}

\begin{figure}
    \centering
    \includegraphics[angle=90,width=1.0\linewidth]{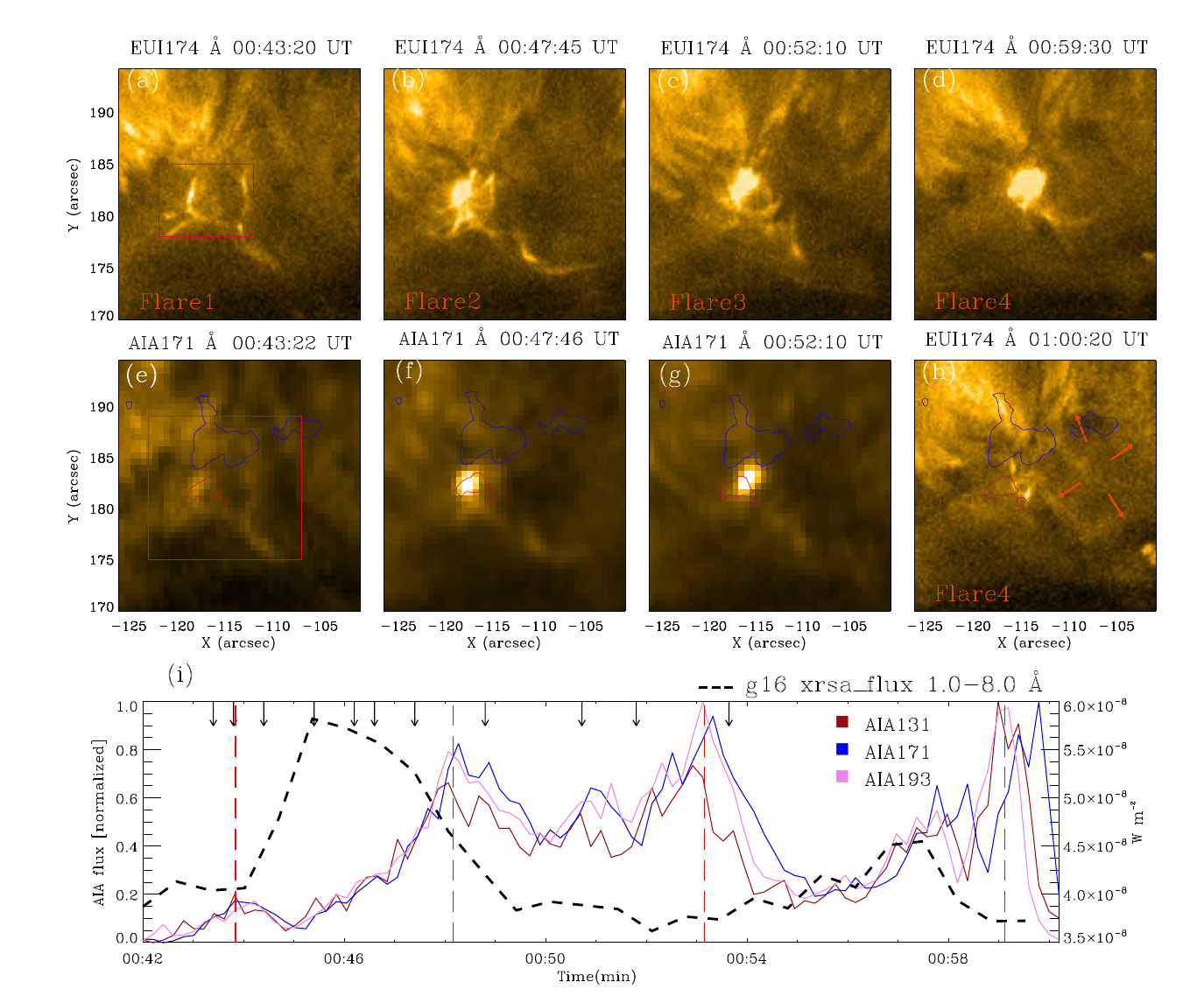} 
       \caption{(a)-(d) and (h) Snapshots of four microflares (Flare 1, Flare 2, Flare 3 and Flare 4) taken in $EUI/HRI_{EUV}$ 174 \AA\ images. (e)-(g) SDO/AIA 171 \AA\ images corresponding in time and space to panels (a), (b) and (c), respectively. The red arrow in panel (h) indicates the transient brightening loop anchored around Flare 4. The contours of SDO/HMI magnetogram are superimposed on panels (e)-(h), with blue and red contours indicating negative and positive polarities, respectively. (i) Normalized flux profiles of the three SDO/AIA EUV channels with in the red box in panel (a). The time series starts at 00:41:56 UT. The black curve represents the GOES X-ray flux. The black arrows indicate the times of the 11 tiny ejections, while red dashed vertical lines represent the peak times of the four flares. An animation of panels (a) and (e) is available, including HMI magnetogram and high-pass filtered EUI 174 \AA\ images. It covers 21 minutes of observations beginning at 00:42:21 UT on 2022 March 8. The video duration is 9 seconds.}
       \label{fig:enter-label1}
   \end{figure}

The flares on the quiet Sun on 2022 March 8, were observed by both SDO/AIA and SO/EUI. The initial flare is referred to as Flare 1, followed by three larger microflares designated as Flare 2, Flare 3, and Flare 4, as illustrated in Figure \ref{fig:enter-label1}. A comparison of EUV observations and SDO/HMI data shows that the flares were located within a dipole magnetic field. Observations from SO/EUI show that Flares 1, 2, and 3 were accompanied by a sequence of moving plasmas originating from the flare's center. Initially, the moving plasma often appeared as dot-like structure, but invariably evolved loop-like eventually (see the animation in Figure \ref{fig:enter-label1} for details). We designate these ejections accompanying the microflares as tiny ejections. In contrast, Flare 4 did not exhibit any discernible tiny ejection.   

\begin{figure}
             \centering
             \includegraphics[width=1.0\linewidth]{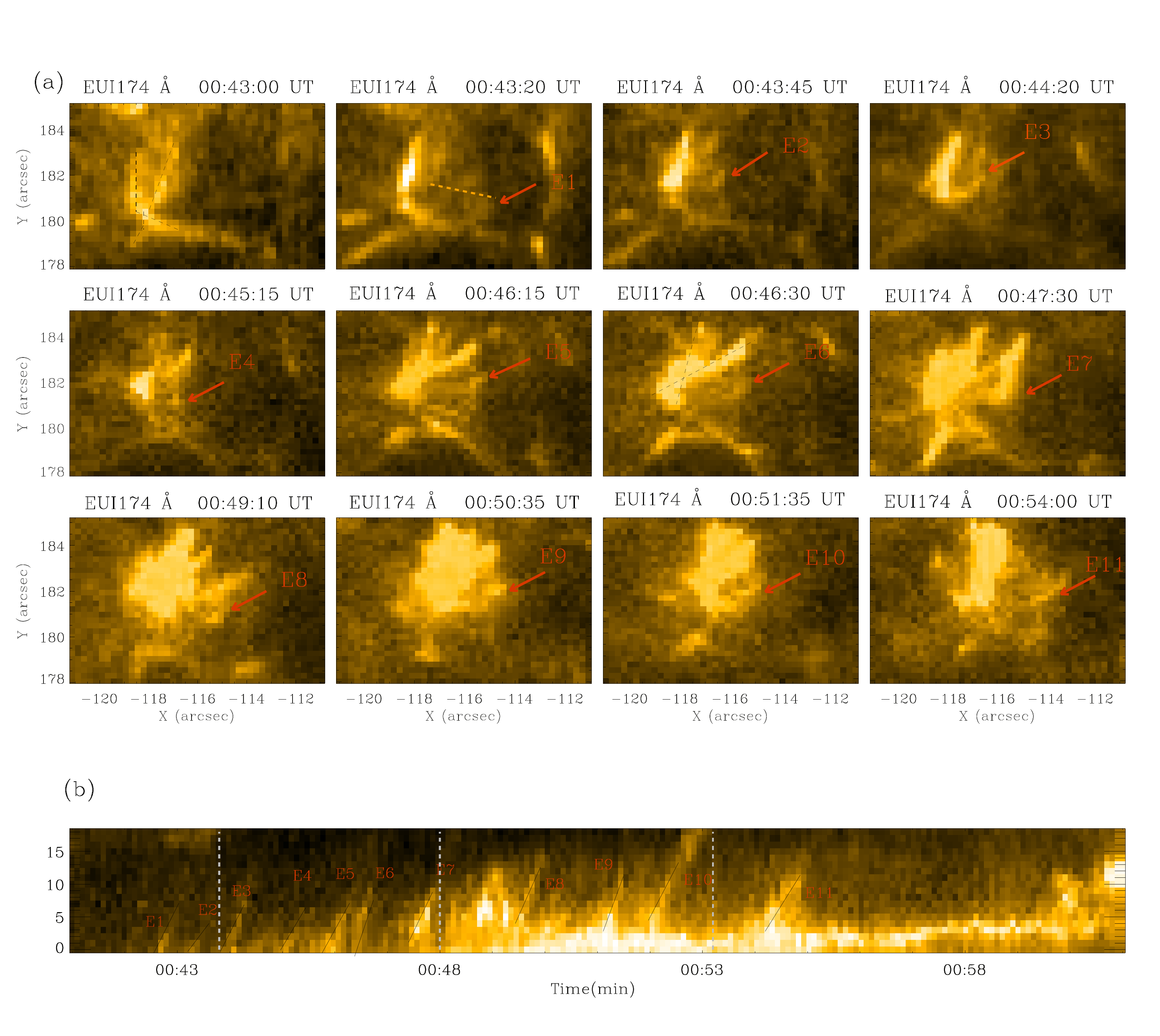}
             \caption{(a) Snapshots of 11 tiny ejections (E1-E11, as marked by red arrows) captured in $EUI/HRI_{EUV}$ 174 \AA\ images. Black dashed lines indicate the crossing flaring loops. (b) Time-distance plots taken along the orange dashed lines plotted on panel (a). White dashed lines represent the peak times of the first three flares,respectively. The plane-of-sky velocities of tiny ejections E1--E11, appearing as inclined bright bands on panel (b), is measured as approximately 232, 154, 229, 216, 221, 339, 248, 261, 317, 226, and 211$km~s^{-1}$, respectively.}
       \label{fig:enter-label2}
   \end{figure}

During Flares 1 and 2, the flaring loops appeared to intersect with each other while roughly aligning along the north-south direction, as indicated by the dashed lines overlaid in Figure \ref{fig:enter-label2}(a). Figure \ref{fig:enter-label2} illustrates eleven tiny ejections, labeled E1-E11, which originated approximately from the crossing points of the initial flaring loops. These ejections propagated predominantly westward with a slight southward component.

A time-slice is used to track the tiny ejections along the orange dashed line shown in Figure \ref{fig:enter-label2}(a). The tiny ejections appear as oblique bright bands in the slice plots, with the oblique lines marking the tiny ejections (E1--E11). The measurement of the slopes of these labeled bands indicates the velocity of the tiny ejections in the plane of the sky ranges from 150 to 340 $km~s^{-1}$. Furthermore, the slices reveal that most tiny ejections have lifetimes of about 20 seconds. In the slice plots, unmarked oblique bands can also be detected during Flares 1 to 3, suggesting the possibility of additional ejections. However, these oblique bands are not observed after Flare 3, consistent with the absence of obvious tiny ejections during Flare 4. 

\begin{figure}
             \centering
             \includegraphics[width=1\linewidth]{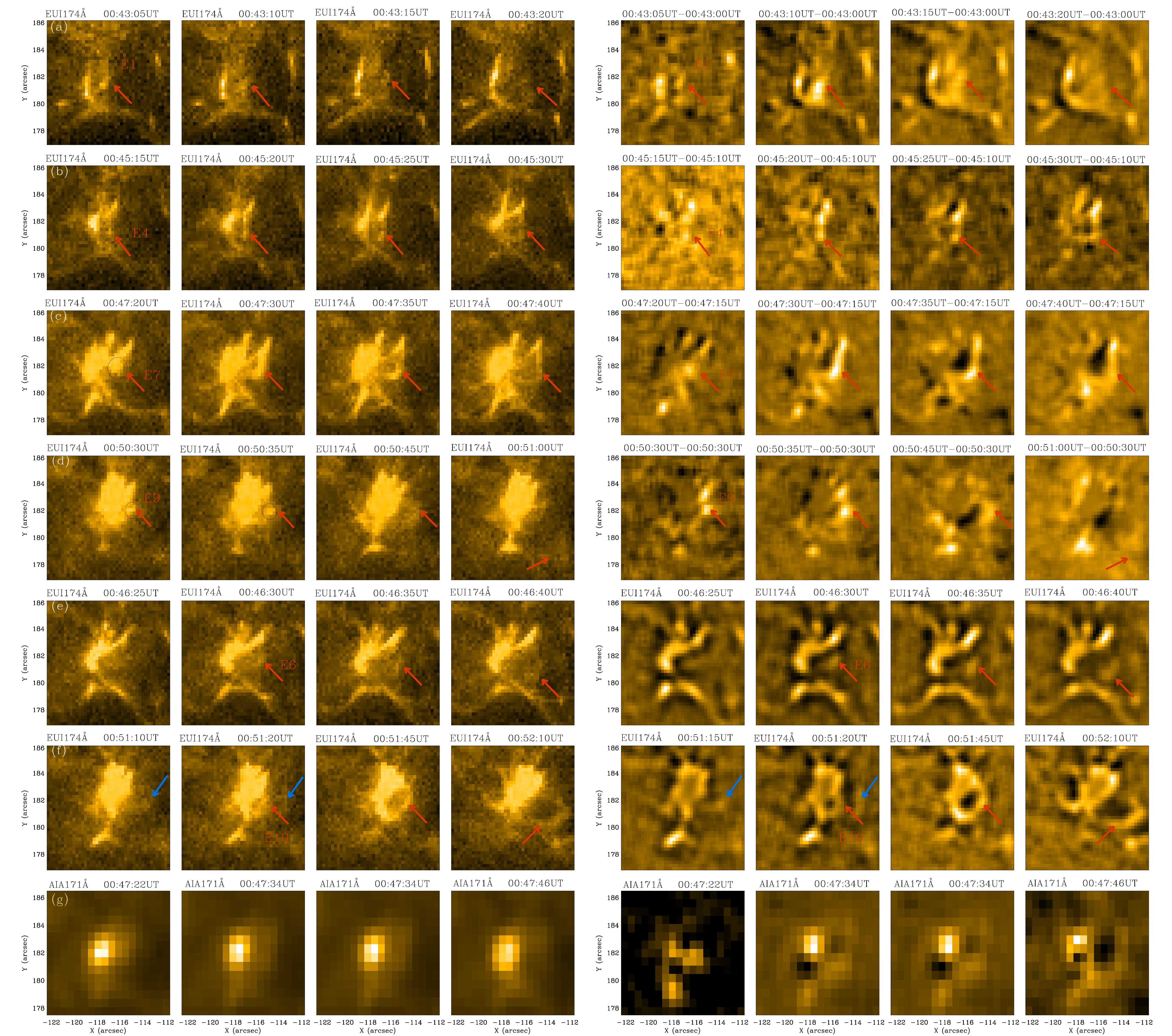}
      \caption{(a)-(f) The $EUI/HRI_{EUV}$ 174 \AA\ images displaying the detailed evolution of several tiny ejections. Red arrows mark tiny ejections, while black circles indicate the initial dot-like structure of tiny ejections. (g) SDO/AIA 171 \AA\ images corresponding to the time and location of panel (c). In each panel, the left four images are raw, while the right four images show their base difference version (a, b, c, d, and g) or high-pass filtered version (e, f).}
       \label{fig:enter-label3}
   \end{figure}

As demonstrated in Figures \ref{fig:enter-label3}(a)-\ref{fig:enter-label3}(d), ejections E1, E4, E7 and E9 are initially appeared in very close proximity to the flare region. In the base-difference EUI images (shown on right side of Figure \ref{fig:enter-label3} (a)-(d)), they are first observed as dot-like structures with length scales ranging from about 3 to 7 pixels, corresponding to 600 to 1200 km. Over time, all of these ejections gradually extended and expanded, eventually evolving into loop-like ones. In contrast, for some ejections, such as E6 and E10, even in the base-difference or high-pass filtering EUI images (right side of Figures \ref{fig:enter-label3}(e) and \ref{fig:enter-label3}(f)), the ejections were not detected as dot-like structures. Instead, they appeared as loop-like formations from the initial observation. Moreover, it is worth noting that these loops exhibit nearly the same alignment as those that evolved from dot-like structures. All of them are approximately parallel to the flaring loops and perpendicular to the polarity inversion line (PIL) of the dipolar region.

Interestingly, the high-pass filtering EUI image in Figure \ref{fig:enter-label3}(f) reveals a dot-like structure nearby the flaring region at 00:51:15 UT, as indicated by the  blue arrows. However, This structure is not associated with the E10 ejection shown in the same figure. After just 5 seconds, it evolved into another distinct loop-like structure in front of the E10 ejection. This observation is consistent with the previous description of the slice analysis and potentially indicates the presence of some smaller, additional tiny ejections.

The observed dot-like structures exhibit a length range of approximately $1.2{\arcsec}$, which corresponds to about 900 km, and is close to the spatial resolution limit of SDO/AIA. This resolution limit explains why these structures cannot be detected in AIA images during the initial formation phase. Even as most tiny ejections evolved into loop structures, their morphological signatures in AIA observations (including E3, E6-E11) were only weakly detected. This limited detection capability prevented AIA from resolving the complete evolutionary sequence from initial dot-like features to fully developed loops. As evidenced in Figure \ref{fig:enter-label3}(g), ejection E7 was clearly captured in EUI and manifested only as faint peripheral loops in AIA images during the flare event.

Flare 4 began at 00:55:21 UT and lasted for 6 minutes. Unlike the first three flares, no tiny ejection was observed during this flare. Instead, $EUI/HRI_{EUV}$ 174 \AA\ images showed that Flare 4 coincided with the formation of a faint, larger-scale transient coronal loop. The loop's southern footpoint was located northwest of the microflare (see red arrows in Figure \ref{fig:enter-label1}(h)).

\begin{figure}
             \centering
             \includegraphics[width=1\linewidth]{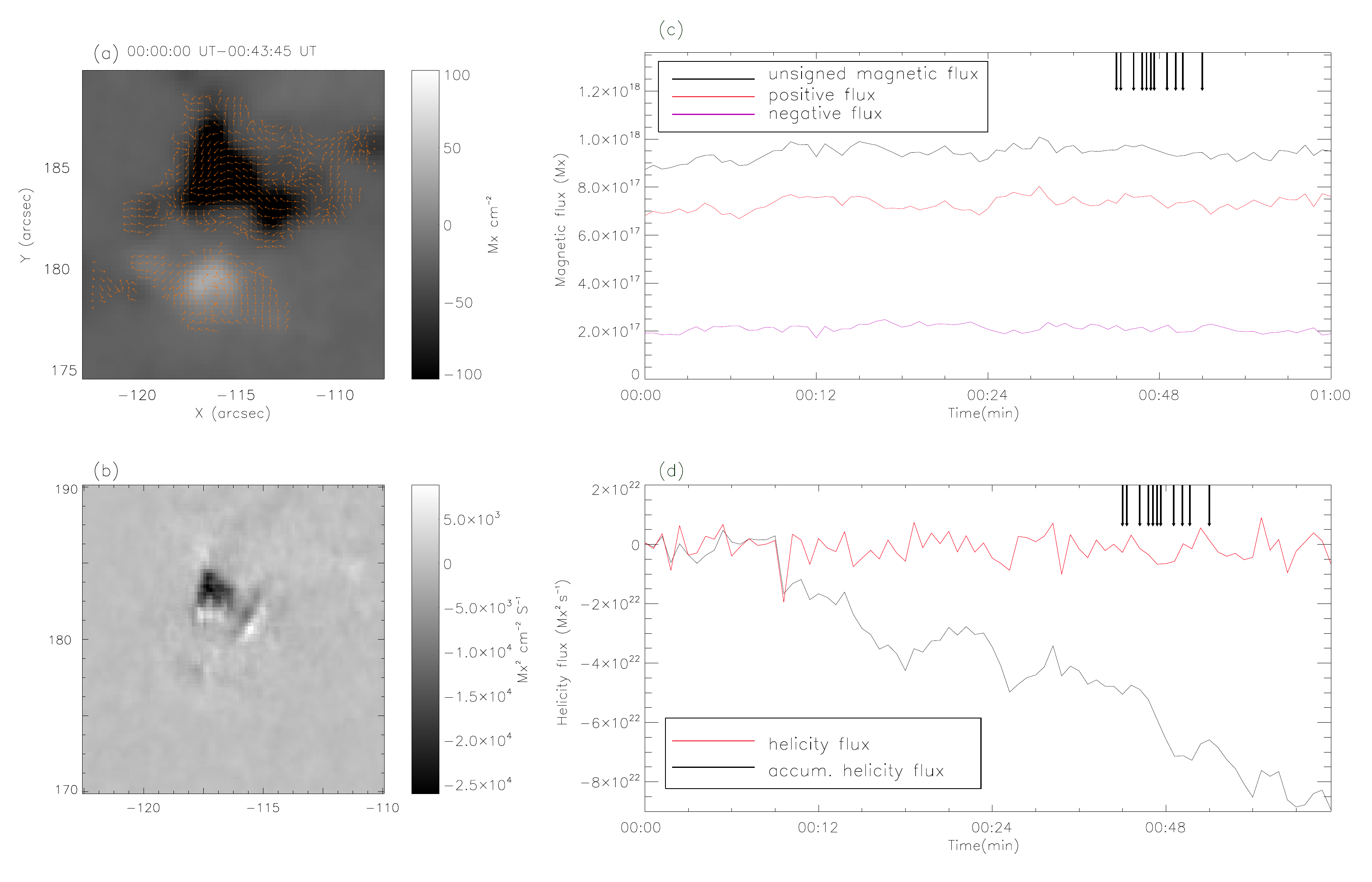}
             \caption{(a) The average SDO/HMI magnetogram from 00:00 UT to 00:43:45 UT within the red box in Figure \ref{fig:enter-label1}(e). The orange arrow indicate the average horizontal velocity field traced from the SDO/HMI magnetograms. (b) The sum of the helicity flux from 00:00 UT to 01:00 UT within region in panel (a). (c) The evolution of HMI magnetograms covering the dipolar field associated with solar flares. The red, violet, and black curves represent the temporal profiles of positive, negative, and unsigned magnetic fluxes, respectively, from 00:00:00 UT to an hour later. The unsigned flux is defined as the summation of positive flux and the absolute value of negative flux. (d) The temporal profile of helicity flux and the accumulated helicity flux. The black and red curves represent the accumulated helicity flux and the helicity flux, respectively, from 00:00:00 UT to one hour later. The black arrows in panels (c) and (d) indicate the times of the 11 tiny ejections. }
       \label{fig:enter-label4}
   \end{figure}

Figure \ref{fig:enter-label4}(c) illustrates the temporal evolution of HMI magnetograms, focusing on the dipolar field associated with solar flares. The red, violet, and black curves represent the temporal profiles of positive, negative, and unsigned magnetic fluxes, respectively, from 00:00:00 UT to an hour later. Throughout this interval, the magnetic flux remained relatively stable, with no significant changes observed. Consequently, there was no evidence of substantial magnetic emergence or cancellation. It is possible that smaller-scale magnetic activities are occurring, the detection of which is currently limited by the spatial resolution and sensitivity thresholds of the HMI instrument. However, as indicated in Figure \ref{fig:enter-label4}(a), the photosphere showed distinct horizontal motion. The velocity field, representing the apparent horizontal movement of magnetic field line footpoints, was derived using the Differential Affine Velocity Estimator (DAVE) method \citep{2006ApJ...646.1358S} applied to HMI magnetograms.

Figure \ref{fig:enter-label4}(d) illustrates the temporal evolution of helicity flux and accumulated helicity flux. The dominant helicity flux was negative, resulting in a significant negative accumulated helicity flux observed before and during the flares. The helicity flux was quantified using the helicity flux density $G_{\theta}$ \citep{2005A&A...439.1191P,2007AdSpR..39.1700C}, calculated from HMI magnetograms in conjunction with the velocity field derived from the DAVE method.

\subsection{Temperature, Energy, and Flow Evolution} 
\begin{figure}
    \centering
    \includegraphics[angle=90,width=0.8\linewidth]{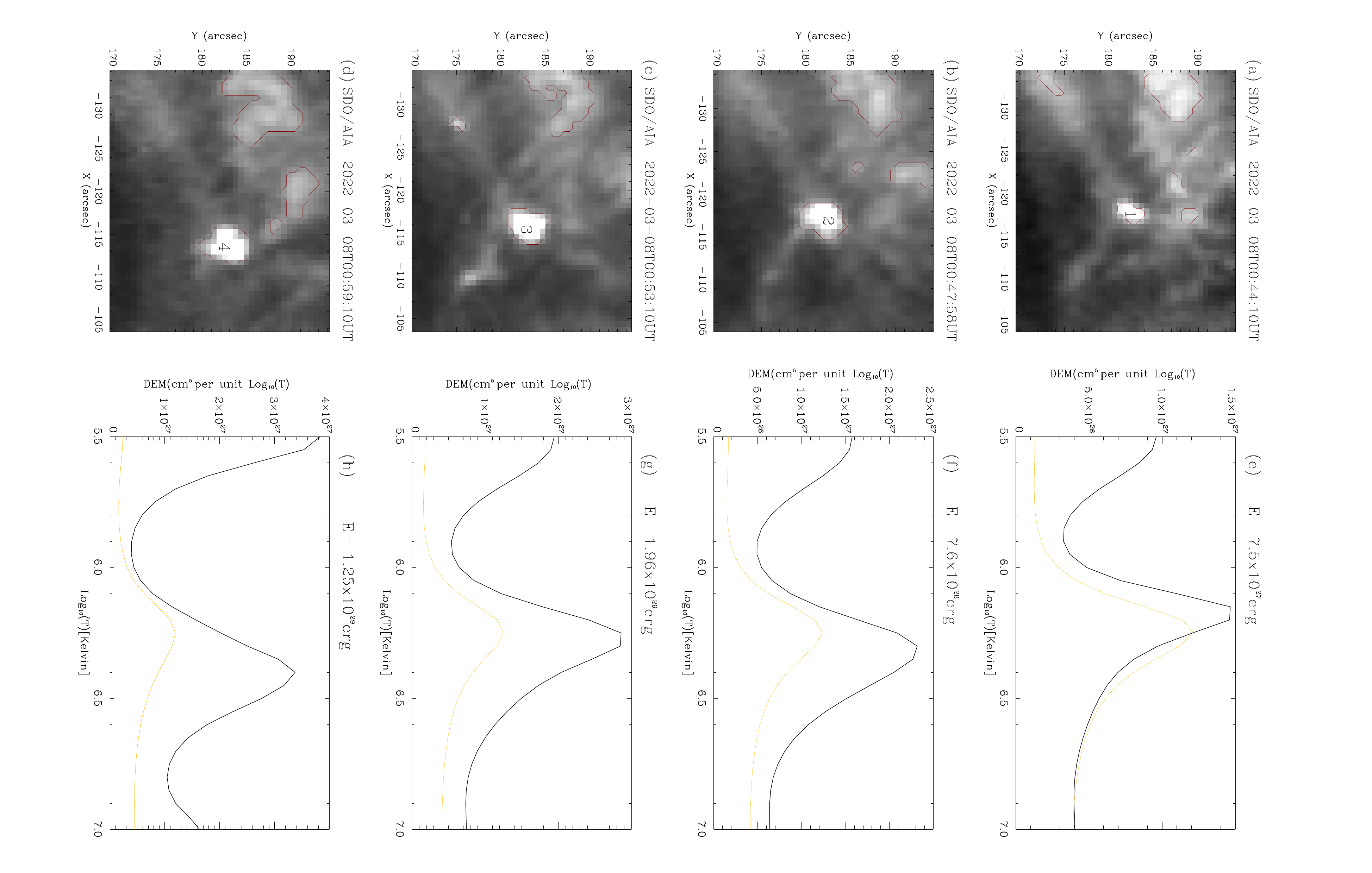}
      \caption{(a)-(d) SDO/AIA 171 \AA\ images at peak times of Flare 1 , Flare 2 , Flare 3 , and Flare 4 , with the regions outlined (1--4). (e)-(h) Profiles of the average DEM temperature within the numbered contours shown on the left panels (a)-(d)}. The yellow curves represent the mean values over a 20-minute period , while the black curve shows the values at peak time of each flare.
       \label{fig:enter-label5}
   \end{figure}

To determine the microflare plasma temperature distribution and energy release, we performed a differential emission measure (DEM) analysis of AIA EUV images using an IDL implementation of the inversion algorithm developed by \cite{2020ApJ...905...17P}. The analysis utilized six aligned AIA EUV channels (94, 131, 171, 193, 211, and 335 \AA). Figure \ref{fig:enter-label5} ((a)-(d)) shows a region of the AIA 171 \AA\ image with intensities above 200 DN, observed at four different times and labeled 1-4, respectively, encompassing the flares under study. The corresponding DEM distributions at each flare's peak time are displayed in the right panels of Figure \ref{fig:enter-label5}. Table \ref{tab:messier} lists the peak DEM temperatures for each flare. Except for Flare 1, Flares 2--4 exhibited peak DEM temperatures at their peak times that exceeded the peak temperature of 20-minute average DEM.

The peak temperature of the DEM for Flare 1 is lower than the background peak temperature primarily because the eruption caused a radiative enhancement in low-temperature bands, such as the 171 \AA\ images, but the region failed to be heated to higher temperatures. Another possibility is that high-temperature bands, like  AIA 94 \AA, did not respond to the weak heating signal, as Flare 1 is one or two order of magnitude weaker than the other flares, as shown below.

Based on the results of the DEM, we estimated the thermal energy released by each flare using the algorithm proposed by \cite{2015ApJ...802...53A}. The results are shown in Table \ref{tab:messier}. Flare 1 released approximately $10^{27}$ erg of thermal energy, an order of magnitude less than Flare 2 and two orders of magnitude less than the ~$10^{29}$ erg observed in Flares 3 and 4. The maximum energy release of the Flares 3 and 4, is also four orders of magnitude smaller than that of the major flares, placing them all in the category of microflares.

We also analyzed the flux evolution in each channels of the AIA to gain further insight into the plasma activity and heating mechanisms of the event. As shown in Figure \ref{fig:enter-label1}(i), the microflare flux profile reveals four main phases of energy release at  131 \AA, 171 \AA, and 193 \AA\ channels. The major peaks in the AIA 131 \AA\ and 193 \AA\ channels appear to be synchronized, while the relative lag of the 171 \AA\ peak reflects the cooling process of the plasma during the microflare's evolution, further supporting the reconnection heating model \citep{2011ApJ...738...24V,2012ApJ...753...35V}. A larger GOES flare occurred during the periods of Flares 1 and 2, but its timing is inconsistent with the AIA flux profile, suggesting it is not directly related. Similarly, the peak time of the second flare detected by GOES does not correspond closely to flare 4 either. Therefore, it should be directed towards the other flares located outside the red box shown in Figure \ref{fig:enter-label1}(a).

\section{Discussion and Conclusions} \label{sec:Conclusions}
 In this letter, we study four successive microflares located in a dipolar field, occurring on 2022 March 8, from 00:42 to 01:02 UT. These microflares occurred at approximately the same location and the previous three microflares were accompanied by multiple tiny ejections. Eleven tiny ejections, clearly visible in EUI/HRI images (but less so in SDO/AIA), were analyzed. These ejections were consistently ejected from the microflare centers and propagated perpendicularly to the flaring loops. Initially, the ejections exhibited dot-like morphologies, often evolving into loop-like structures. 
 
Eruptive EUV loops are often considered to trace eruptive magnetic flux ropes \citep{2012NatCo...3..747Z,2010ApJ...716L..68C,2024ApJ...976..207C,2024ApJ...967..130L}. As observed by \citet{1999SoPh..188..365W}, a microflare was accompanied by a coronal loop eruption from a highly sheared magnetic field, with the loop orientated parallel to the magnetic neutral line. However, our observations present several inconsistencies with this interpretation. Firstly, the occurrence of about a dozen homologous loop eruptions during our event, within a period of about 10 minutes, might suggest the presence of multiple magnetic flux ropes in the dipolar region. The frequency and rapidity of the observed ejections are unusual for typical loop eruptions. Secondly, the loops in our study, regardless of whether they evolved from dot-like structures or not, are consistently observed to be perpendicular to the PIL of the dipolar region. If these loops originated from flux ropes aligned along the PIL, they would have needed to rotate by $90^{\circ}$ upon becoming visible. Such phenomena do not appear to be common in current observations.

Similar to the nanojets reported by \cite{2021NatAs...5...54A}, the tiny ejections in our observations are expelled perpendicular to the coronal loops. This similarity suggests that these ejections may result from magnetic reconnection involving small-angle misalignments of magnetic field lines in braided magnetic fields. However, unlike nanojets, which typically appear as collimated plasma ejections, the tiny ejections in our observations initially manifest as dot-like structures. As they propagate perpendicular to the flaring loops, they travel across the surrounding magnetic fields that are nearly parallel to the flaring loops. Consequently, they inevitably interact with and heat the ambient magnetic field, leading to their evolution from a dot-like shape into a loop-like structure. This transformation provides a plausible explanation for the loop-like eruptions observed in the later stages of the tiny ejections.

The flaring loops crossing with each other were clearly observed during Flare 1 and at the onset of Flare 2. The presence of the crossing coronal loops is often considered evidence of the braided magnetic field in the solar corona \citep{2013Natur.493..501C,2022A&A...667A.166C,2023A&A...679A...9B}. However, the braiding of a magnetic field typically implies field lines that are closely entangled, with the formation of current sheets between the loop components of the braid. These loops could be rooted in distant magnetic polarities and may cross ``by chance" from the specific observational perspective, potentially remaining far from each other in reality. As noted by \cite{2017ApJ...837..108P}, the braiding magnetic reconnection can only occur when the magnetic loops interact with each other. However, the observed loop structures in this study provide crucial clues about the reconnection sites. The continual tiny ejections were observed to originate at the crossing points of the flaring loops, strongly suggesting that magnetic reconnection may be occurring in this region, where the magnetic field would be significantly misaligned. Nevertheless, it's essential to note that the magnetic fields participating in the reconnection process are not necessarily precisely outlined by the visible crossing loops. The actual magnetic field dynamics may be more complex than what can be directly observed in the EUV images.

The preceding reconnections among the braiding magnetic field may facilitate further braiding reconnection events, potentially triggering subsequent flares \citep{1976GAM.....8.....S,1988ApJ...330..474P,2018A&A...615A..84R,2024A&A...689A.184C}. The observed sequence of flares in our study, occurring in rapid succession, aligns with this theoretical framework, although a causal relationship between individual flares cannot be definitively established based on the available data.

We also investigated the horizontally moving plasma flow associated with the dipolar regions where the microflares occurred. The organized horizontal motion of magnetic flux in the photosphere, particularly shearing and rotating motions, can inject magnetic helicity into the corona by twisting the coronal field \citep{2003SoPh..215..203D,Liu_2012,2015ApJ...805...48B}. Continuous negative helicity flux was observed around the studied region approximately one hour before the microflares. Such helicity accumulation is expected to drive progressive twisting of the coronal dipole field. In fact, a slightly twisted dipole magnetic field may favour the creation of such small-angle magnetic reconnections.

 During Flare 4, however, no similar tiny ejections were observed. Instead, the microflare accompanied the brightening of a large-scale loop  with one footpoint rooted in the dipolar fields. Therefore, it seems that the flare was produced by magnetic reconnection between the dipolar field and the surrounding field. As discussed earlier, the dipolar field  was twisted to some extent, and a series of small-angle reconnections may have occurred within it during previous three microflares. According to simulations by \cite{2016ApJ...817....5H}, some twisted magnetic field lines remain after a series of reconnections within the braided field. Additionally, the reconnection of braided structures can lead to the expansion of non-potential magnetic structures \citep{2018A&A...615A..84R}. Therefore, the expansion of non-potential magnetic structures could trigger interactions between the non-potential field and the surrounding field. This scenario may be similar to that observed in usual jets or blowout jets.

Our study reveals some physical processes that are different from previous microflare models. The main results and summaries are as follows.
\begin{itemize}
\item[1] The first three successive microflares, out of the four observed, were accompanied by  small-scale moving plasma structures, which we termed as  tiny ejections. They are generated near the center of the flaring loops.
\item[2]These tiny ejections propagated approximately perpendicular to the initial flaring loops. Initially, they often appeared as dot-like structures, but consistently evolved into loop-like shapes over time.
\item[3]We propose that these tiny ejections result from magnetic reconnection caused by the small-angle misalignment of braided magnetic field lines. This proposal mechanism may be responsible for triggering the microflares associated with the tiny ejections.
\end{itemize} 

A question that extends from this paper is how frequently small-angle reconnections within braided magnetic fields lead to microflare eruptions and whether the type of magnetic reconnection implied by the tiny ejections could drive larger flare eruptions. 

%% Also note that the akcnowlodgment environment does not support long amounts of text. If you have a lot of people and institutions to acknowledge, do not use this command. Instead, create a new 



\section*{Acknowledgements} 
We thank the referee, who raised valuable comments to improve the manuscript. This work was supported by the National Science Foundation of China under grants12273106 and 12273108, and the CAS “Light of West China" Program, and the "Yunnan Revitalization Talent Support Program " Innovation Team Project (202405AS 350012). SO is a space mission of international collaboration between ESA and NASA, operated by ESA. The EUI instrument was built by CSL, IAS, MPS, MSSL/UCL, PMOD/WRC, ROB, LCF/IO with funding from the Belgian Federal Science Policy Office (BELSPO/PRODEX PEA 4000112292); the Centre National d’Etudes Spatiales (CNES); the UK Space Agency (UKSA); the Bundesministerium für Wirtschaft und Energie (BMWi) through the Deutsches Zentrum für Luft-und Raumfahrt (DLR); and the Swiss Space Office (SSO). AIA and HMI are instruments on board the Solar Dynamics Observatory, a mission for NASA’s Living With a Star program.

\bibliography{ms.bib}

\begin{thebibliography}{}
\expandafter\ifx\csname natexlab\endcsname\relax\def\natexlab#1{#1}\fi
\providecommand{\url}[1]{\href{#1}{#1}}
\providecommand{\dodoi}[1]{doi:~\href{http://doi.org/#1}{\nolinkurl{#1}}}
\providecommand{\doeprint}[1]{\href{http://ascl.net/#1}{\nolinkurl{http://ascl.net/#1}}}
\providecommand{\doarXiv}[1]{\href{https://arxiv.org/abs/#1}{\nolinkurl{https://arxiv.org/abs/#1}}}

\bibitem[{P. {Antolin} {et~al.}(2021){Antolin}, {Pagano}, {Testa}, {Petralia},
  \& {Reale}}]{2021NatAs...5...54A}
{Antolin}, P., {Pagano}, P., {Testa}, P., {Petralia}, A., \& {Reale}, F. 2021,
  \bibinfo{title}{{Reconnection nanojets in the solar corona},} Nature
  Astronomy, 5, 54, \dodoi{10.1038/s41550-020-1199-8}

\bibitem[{M.~J. {Aschwanden}(2022){Aschwanden}}]{2022ApJ...934L...3A}
{Aschwanden}, M.~J. 2022, \bibinfo{title}{{Reconciling Power-law Slopes in
  Solar Flare and Nanoflare Size Distributions},} \apjl, 934, L3,
  \dodoi{10.3847/2041-8213/ac7b8d}

\bibitem[{M.~J. {Aschwanden} {et~al.}(2015){Aschwanden}, {Boerner}, {Ryan},
  {Caspi}, {McTiernan}, \& {Warren}}]{2015ApJ...802...53A}
{Aschwanden}, M.~J., {Boerner}, P., {Ryan}, D., {et~al.} 2015,
  \bibinfo{title}{{Global Energetics of Solar Flares: II. Thermal Energies},}
  \apj, 802, 53, \dodoi{10.1088/0004-637X/802/1/53}

\bibitem[{M.~J. {Aschwanden} \& C.~E. {Parnell}(2002){Aschwanden} \&
  {Parnell}}]{2002ApJ...572.1048A}
{Aschwanden}, M.~J., \& {Parnell}, C.~E. 2002, \bibinfo{title}{{Nanoflare
  Statistics from First Principles: Fractal Geometry and Temperature
  Synthesis},} \apj, 572, 1048, \dodoi{10.1086/340385}

\bibitem[{A.~F. {Battaglia} {et~al.}(2023){Battaglia}, {Wang}, {Saqri},
  {Podladchikova}, {Veronig}, {Collier}, {Dickson}, {Podladchikova},
  {Monstein}, {Warmuth}, {Schuller}, {Harra}, \&
  {Krucker}}]{2023A&A...670A..56B}
{Battaglia}, A.~F., {Wang}, W., {Saqri}, J., {et~al.} 2023,
  \bibinfo{title}{{Identifying the energy release site in a solar microflare
  with a jet},} \aap, 670, A56, \dodoi{10.1051/0004-6361/202244996}

\bibitem[{D. {Berghmans} {et~al.}(2021){Berghmans}, {Auch{\`e}re}, {Long},
  {Soubri{\'e}}, {Mierla}, {Zhukov}, {Sch{\"u}hle}, {Antolin}, {Harra},
  {Parenti}, {Podladchikova}, {Aznar Cuadrado}, {Buchlin}, {Dolla}, {Verbeeck},
  {Gissot}, {Teriaca}, {Haberreiter}, {Katsiyannis}, {Rodriguez}, {Kraaikamp},
  {Smith}, {Stegen}, {Rochus}, {Halain}, {Jacques}, {Thompson}, \&
  {Inhester}}]{2021A&A...656L...4B}
{Berghmans}, D., {Auch{\`e}re}, F., {Long}, D.~M., {et~al.} 2021,
  \bibinfo{title}{{Extreme-UV quiet Sun brightenings observed by the Solar
  Orbiter/EUI},} \aap, 656, L4, \dodoi{10.1051/0004-6361/202140380}

\bibitem[{D. {Berghmans} {et~al.}(2023){Berghmans}, {Antolin}, {Auch{\`e}re},
  {Aznar Cuadrado}, {Barczynski}, {Chitta}, {Gissot}, {Harra}, {Huang},
  {Janvier}, {Kraaikamp}, {Long}, {Mandal}, {Mierla}, {Parenti}, {Peter},
  {Rodriguez}, {Sch{\"u}hle}, {Smith}, {Solanki}, {Stegen}, {Teriaca},
  {Verbeeck}, {West}, {Zhukov}, {Appourchaux}, {Aulanier}, {Buchlin},
  {Delmotte}, {Gilles}, {Haberreiter}, {Halain}, {Heerlein}, {Hochedez}, {Gyo},
  {Poedts}, {Renotte}, \& {Rochus}}]{2023A&A...675A.110B}
{Berghmans}, D., {Antolin}, P., {Auch{\`e}re}, F., {et~al.} 2023,
  \bibinfo{title}{{First perihelion of EUI on the Solar Orbiter mission},}
  \aap, 675, A110, \dodoi{10.1051/0004-6361/202245586}

\bibitem[{Y. {Bi} {et~al.}(2015){Bi}, {Jiang}, {Yang}, {Xiang}, {Cai}, \&
  {Liu}}]{2015ApJ...805...48B}
{Bi}, Y., {Jiang}, Y., {Yang}, J., {et~al.} 2015, \bibinfo{title}{{Partial
  Eruption of a Filament with Twisting Non-uniform Fields},} \apj, 805, 48,
  \dodoi{10.1088/0004-637X/805/1/48}

\bibitem[{Y. {Bi} {et~al.}(2023){Bi}, {Yang}, {Qin}, {Qiang}, {Hong}, {Yang},
  {Xu}, {Liu}, \& {Ji}}]{2023A&A...679A...9B}
{Bi}, Y., {Yang}, J.-Y., {Qin}, Y., {et~al.} 2023,
  \bibinfo{title}{{Morphological evidence for nanoflares heating warm loops in
  the solar corona},} \aap, 679, A9, \dodoi{10.1051/0004-6361/202346944}

\bibitem[{J.~W. {Brosius} \& G.~D. {Holman}(2009){Brosius} \&
  {Holman}}]{2009ApJ...692..492B}
{Brosius}, J.~W., \& {Holman}, G.~D. 2009, \bibinfo{title}{{Observations of the
  Thermal and Dynamic Evolution of a Solar Microflare},} \apj, 692, 492,
  \dodoi{10.1088/0004-637X/692/1/492}

\bibitem[{P.~K. {Browning} {et~al.}(2008){Browning}, {Gerrard}, {Hood},
  {Kevis}, \& {van der Linden}}]{2008A&A...485..837B}
{Browning}, P.~K., {Gerrard}, C., {Hood}, A.~W., {Kevis}, R., \& {van der
  Linden}, R.~A.~M. 2008, \bibinfo{title}{{Heating the corona by nanoflares:
  simulations of energy release triggered by a kink instability},} \aap, 485,
  837, \dodoi{10.1051/0004-6361:20079192}

\bibitem[{J. {Chae}(2007){Chae}}]{2007AdSpR..39.1700C}
{Chae}, J. 2007, \bibinfo{title}{{Measurements of magnetic helicity injected
  through the solar photosphere},} Advances in Space Research, 39, 1700,
  \dodoi{10.1016/j.asr.2007.01.035}

\bibitem[{F. {Chen} \& M.~D. {Ding}(2010){Chen} \&
  {Ding}}]{2010ApJ...724..640C}
{Chen}, F., \& {Ding}, M.~D. 2010, \bibinfo{title}{{Evidence of Explosive
  Evaporation in a Microflare Observed by Hinode/EIS},} \apj, 724, 640,
  \dodoi{10.1088/0004-637X/724/1/640}

\bibitem[{H. {Chen} {et~al.}(2017){Chen}, {Zhang}, {Ma}, {Yan}, \&
  {Xue}}]{2017ApJ...841L..13C}
{Chen}, H., {Zhang}, J., {Ma}, S., {Yan}, X., \& {Xue}, J. 2017,
  \bibinfo{title}{{Solar Tornadoes Triggered by Interaction between Filaments
  and EUV Jets},} \apjl, 841, L13, \dodoi{10.3847/2041-8213/aa71a2}

\bibitem[{H. {Chen} {et~al.}(2024){Chen}, {Fletcher}, {Zhou}, {Cheng}, {Wang},
  {Mulay}, {Zheng}, {Ma}, \& {Zhang}}]{2024ApJ...976..207C}
{Chen}, H., {Fletcher}, L., {Zhou}, G., {et~al.} 2024,
  \bibinfo{title}{{Simultaneous Eruption and Shrinkage of Preexisting Flare
  Loops during a Subsequent Solar Eruption},} \apj, 976, 207,
  \dodoi{10.3847/1538-4357/ad8c25}

\bibitem[{P.~F. Chen {et~al.}(1999)Chen, Fang, Ding, \& Tang}]{Chen_1999}
Chen, P.~F., Fang, C., Ding, M.~D., \& Tang, Y.~H. 1999,
  \bibinfo{title}{Flaring Loop Motion and a Unified Model for Solar Flares,}
  The Astrophysical Journal, 520, 853, \dodoi{10.1086/307477}

\bibitem[{X. {Cheng} {et~al.}(2010){Cheng}, {Ding}, {Guo}, {Zhang}, {Jing}, \&
  {Wiegelmann}}]{2010ApJ...716L..68C}
{Cheng}, X., {Ding}, M.~D., {Guo}, Y., {et~al.} 2010,
  \bibinfo{title}{{Re-flaring of a Post-flare Loop System Driven by Flux Rope
  Emergence and Twisting},} \apjl, 716, L68,
  \dodoi{10.1088/2041-8205/716/1/L68}

\bibitem[{C. {Chifor} {et~al.}(2008){Chifor}, {Isobe}, {Mason}, {Hannah},
  {Young}, {Del Zanna}, {Krucker}, {Ichimoto}, {Katsukawa}, \&
  {Yokoyama}}]{2008A&A...491..279C}
{Chifor}, C., {Isobe}, H., {Mason}, H.~E., {et~al.} 2008,
  \bibinfo{title}{{Magnetic flux cancellation associated with a recurring solar
  jet observed with Hinode, RHESSI, and STEREO/EUVI},} \aap, 491, 279,
  \dodoi{10.1051/0004-6361:200810265}

\bibitem[{L.~P. {Chitta} {et~al.}(2022){Chitta}, {Peter}, {Parenti},
  {Berghmans}, {Auch{\`e}re}, {Solanki}, {Aznar Cuadrado}, {Sch{\"u}hle},
  {Teriaca}, {Mandal}, {Barczynski}, {Buchlin}, {Harra}, {Kraaikamp}, {Long},
  {Rodriguez}, {Schwanitz}, {Smith}, {Verbeeck}, {Zhukov}, {Liu}, \&
  {Cheung}}]{2022A&A...667A.166C}
{Chitta}, L.~P., {Peter}, H., {Parenti}, S., {et~al.} 2022,
  \bibinfo{title}{{Solar coronal heating from small-scale magnetic braids},}
  \aap, 667, A166, \dodoi{10.1051/0004-6361/202244170}

\bibitem[{S. {Christe} {et~al.}(2008){Christe}, {Hannah}, {Krucker},
  {McTiernan}, \& {Lin}}]{2008ApJ...677.1385C}
{Christe}, S., {Hannah}, I.~G., {Krucker}, S., {McTiernan}, J., \& {Lin}, R.~P.
  2008, \bibinfo{title}{{RHESSI Microflare Statistics. I. Flare-Finding and
  Frequency Distributions},} \apj, 677, 1385, \dodoi{10.1086/529011}

\bibitem[{J.~W. {Cirtain} {et~al.}(2013){Cirtain}, {Golub}, {Winebarger}, {de
  Pontieu}, {Kobayashi}, {Moore}, {Walsh}, {Korreck}, {Weber}, {McCauley},
  {Title}, {Kuzin}, \& {Deforest}}]{2013Natur.493..501C}
{Cirtain}, J.~W., {Golub}, L., {Winebarger}, A.~R., {et~al.} 2013,
  \bibinfo{title}{{Energy release in the solar corona from spatially resolved
  magnetic braids},} \nat, 493, 501, \dodoi{10.1038/nature11772}

\bibitem[{G. {Cozzo} {et~al.}(2024){Cozzo}, {Reid}, {Pagano}, {Reale}, {Testa},
  {Hood}, {Argiroffi}, {Petralia}, {Alaimo}, {D'Anca}, {Sciortino}, {Todaro},
  {Lo Cicero}, {Barbera}, {de Pontieu}, \&
  {Martinez-Sykora}}]{2024A&A...689A.184C}
{Cozzo}, G., {Reid}, J., {Pagano}, P., {et~al.} 2024, \bibinfo{title}{{Coronal
  energy release by MHD avalanches: II. EUV line emission from a multi-threaded
  coronal loop},} \aap, 689, A184, \dodoi{10.1051/0004-6361/202450644}

\bibitem[{P. {D{\'e}moulin} \& M.~A. {Berger}(2003){D{\'e}moulin} \&
  {Berger}}]{2003SoPh..215..203D}
{D{\'e}moulin}, P., \& {Berger}, M.~A. 2003, \bibinfo{title}{{Magnetic Energy
  and Helicity Fluxes at the Photospheric Level},} \solphys, 215, 203,
  \dodoi{10.1023/A:1025679813955}

\bibitem[{U. {Feldman} {et~al.}(1996){Feldman}, {Doschek}, {Behring}, \&
  {Phillips}}]{1996ApJ...460.1034F}
{Feldman}, U., {Doschek}, G.~A., {Behring}, W.~E., \& {Phillips}, K.~J.~H.
  1996, \bibinfo{title}{{Electron Temperature, Emission Measure, and X-Ray Flux
  in A2 to X2 X-Ray Class Solar Flares},} \apj, 460, 1034,
  \dodoi{10.1086/177030}

\bibitem[{L. {Glesener} {et~al.}(2017){Glesener}, {Krucker}, {Hannah},
  {Hudson}, {Grefenstette}, {White}, {Smith}, \& {Marsh}}]{2017ApJ...845..122G}
{Glesener}, L., {Krucker}, S., {Hannah}, I.~G., {et~al.} 2017,
  \bibinfo{title}{{NuSTAR Hard X-Ray Observation of a Sub-A Class Solar
  Flare},} \apj, 845, 122, \dodoi{10.3847/1538-4357/aa80e9}

\bibitem[{L. {Glesener} {et~al.}(2020){Glesener}, {Krucker}, {Duncan},
  {Hannah}, {Grefenstette}, {Chen}, {Smith}, {White}, \&
  {Hudson}}]{2020ApJ...891L..34G}
{Glesener}, L., {Krucker}, S., {Duncan}, J., {et~al.} 2020,
  \bibinfo{title}{{Accelerated Electrons Observed Down to <7 keV in a NuSTAR
  Solar Microflare},} \apjl, 891, L34, \dodoi{10.3847/2041-8213/ab7341}

\bibitem[{I.~G. {Hannah} {et~al.}(2011){Hannah}, {Hudson}, {Battaglia},
  {Christe}, {Ka{\v{s}}parov{\'a}}, {Krucker}, {Kundu}, \&
  {Veronig}}]{2011SSRv..159..263H}
{Hannah}, I.~G., {Hudson}, H.~S., {Battaglia}, M., {et~al.} 2011,
  \bibinfo{title}{{Microflares and the Statistics of X-ray Flares},} \ssr, 159,
  263, \dodoi{10.1007/s11214-010-9705-4}

\bibitem[{J. {Heyvaerts} {et~al.}(1977){Heyvaerts}, {Priest}, \&
  {Rust}}]{1977ApJ...216..123H}
{Heyvaerts}, J., {Priest}, E.~R., \& {Rust}, D.~M. 1977, \bibinfo{title}{{An
  emerging flux model for the solar phenomenon.},} \apj, 216, 123,
  \dodoi{10.1086/155453}

\bibitem[{A.~W. {Hood} {et~al.}(2016){Hood}, {Cargill}, {Browning}, \&
  {Tam}}]{2016ApJ...817....5H}
{Hood}, A.~W., {Cargill}, P.~J., {Browning}, P.~K., \& {Tam}, K.~V. 2016,
  \bibinfo{title}{{An MHD Avalanche in a Multi-threaded Coronal Loop.},} \apj,
  817, 5, \dodoi{10.3847/0004-637X/817/1/5}

\bibitem[{Z. {Hou} {et~al.}(2021){Hou}, {Tian}, {Berghmans}, {Chen}, {Teriaca},
  {Sch{\"u}hle}, {Gao}, {Chen}, {He}, {Wang}, \& {Bai}}]{2021ApJ...918L..20H}
{Hou}, Z., {Tian}, H., {Berghmans}, D., {et~al.} 2021, \bibinfo{title}{{Coronal
  Microjets in Quiet-Sun Regions Observed with the Extreme Ultraviolet Imager
  on Board the Solar Orbiter},} \apjl, 918, L20,
  \dodoi{10.3847/2041-8213/ac1f30}

\bibitem[{H.~S. {Hudson}(1991{\natexlab{a}}){Hudson}}]{1991SoPh..133..357H}
{Hudson}, H.~S. 1991{\natexlab{a}}, \bibinfo{title}{{Solar flares, microflares,
  nanoflares, and coronal heating},} \solphys, 133, 357,
  \dodoi{10.1007/BF00149894}

\bibitem[{H.~S. {Hudson}(1991{\natexlab{b}}){Hudson}}]{1991BAAS...23R1064H}
{Hudson}, H.~S. 1991{\natexlab{b}}, in Bulletin of the American Astronomical
  Society, Vol.~23, 1064

\bibitem[{A.~R. Inglis \& S. Christe(2014)Inglis \& Christe}]{Inglis_2014}
Inglis, A.~R., \& Christe, S. 2014, \bibinfo{title}{INVESTIGATING THE
  DIFFERENTIAL EMISSION MEASURE AND ENERGETICS OF MICROFLARES WITH COMBINED
  SDO/AIA AND RHESSI OBSERVATIONS,} The Astrophysical Journal, 789, 116,
  \dodoi{10.1088/0004-637X/789/2/116}

\bibitem[{D.~B. {Jess} {et~al.}(2010){Jess}, {Mathioudakis}, {Browning},
  {Crockett}, \& {Keenan}}]{2010ApJ...712L.111J}
{Jess}, D.~B., {Mathioudakis}, M., {Browning}, P.~K., {Crockett}, P.~J., \&
  {Keenan}, F.~P. 2010, \bibinfo{title}{{Microflare Activity Driven by Forced
  Magnetic Reconnection},} \apjl, 712, L111,
  \dodoi{10.1088/2041-8205/712/1/L111}

\bibitem[{F. {Jiang} {et~al.}(2015){Jiang}, {Zhang}, \&
  {Yang}}]{2015PASJ...67...40J}
{Jiang}, F., {Zhang}, J., \& {Yang}, S. 2015, \bibinfo{title}{{Relationship
  between extreme ultraviolet microflares and small-scale magnetic fields in
  the quiet Sun},} \pasj, 67, 40, \dodoi{10.1093/pasj/psv009}

\bibitem[{R. {Kano} {et~al.}(2010){Kano}, {Shimizu}, \&
  {Tarbell}}]{2010ApJ...720.1136K}
{Kano}, R., {Shimizu}, T., \& {Tarbell}, T.~D. 2010, \bibinfo{title}{{Hinode
  Observation of Photospheric Magnetic Activities Triggering X-ray Microflares
  Around a Well-developed Sunspot},} \apj, 720, 1136,
  \dodoi{10.1088/0004-637X/720/2/1136}

\bibitem[{J.~R. {Lemen} {et~al.}(2012){Lemen}, {Title}, {Akin}, {Boerner},
  {Chou}, {Drake}, {Duncan}, {Edwards}, {Friedlaender}, {Heyman}, {Hurlburt},
  {Katz}, {Kushner}, {Levay}, {Lindgren}, {Mathur}, {McFeaters}, {Mitchell},
  {Rehse}, {Schrijver}, {Springer}, {Stern}, {Tarbell}, {Wuelser}, {Wolfson},
  {Yanari}, {Bookbinder}, {Cheimets}, {Caldwell}, {Deluca}, {Gates}, {Golub},
  {Park}, {Podgorski}, {Bush}, {Scherrer}, {Gummin}, {Smith}, {Auker},
  {Jerram}, {Pool}, {Soufli}, {Windt}, {Beardsley}, {Clapp}, {Lang}, \&
  {Waltham}}]{2012SoPh..275...17L}
{Lemen}, J.~R., {Title}, A.~M., {Akin}, D.~J., {et~al.} 2012,
  \bibinfo{title}{{The Atmospheric Imaging Assembly (AIA) on the Solar Dynamics
  Observatory (SDO)},} \solphys, 275, 17, \dodoi{10.1007/s11207-011-9776-8}

\bibitem[{L. {Li} {et~al.}(2024){Li}, {Song}, {Peter}, {Chitta}, {Cheng}, {Li},
  \& {Zhou}}]{2024ApJ...967..130L}
{Li}, L., {Song}, H., {Peter}, H., {et~al.} 2024, \bibinfo{title}{{Eruption of
  a Million-Kelvin Warm Magnetic Flux Rope on the Sun},} \apj, 967, 130,
  \dodoi{10.3847/1538-4357/ad3fb3}

\bibitem[{Z.~F. {Li} {et~al.}(2022){Li}, {Cheng}, {Chen}, {Chen}, \&
  {Ding}}]{2022ApJ...930L...7L}
{Li}, Z.~F., {Cheng}, X., {Chen}, F., {Chen}, J., \& {Ding}, M.~D. 2022,
  \bibinfo{title}{{Three-dimensional Magnetic and Thermodynamic Structures of
  Solar Microflares},} \apjl, 930, L7, \dodoi{10.3847/2041-8213/ac67aa}

\bibitem[{Y. Liu \& P.~W. Schuck(2012)Liu \& Schuck}]{Liu_2012}
Liu, Y., \& Schuck, P.~W. 2012, \bibinfo{title}{MAGNETIC ENERGY AND HELICITY IN
  TWO EMERGING ACTIVE REGIONS IN THE SUN,} The Astrophysical Journal, 761, 105,
  \dodoi{10.1088/0004-637X/761/2/105}

\bibitem[{R.~L. Moore {et~al.}(2010)Moore, Cirtain, Sterling, \&
  Falconer}]{Moore_2010}
Moore, R.~L., Cirtain, J.~W., Sterling, A.~C., \& Falconer, D.~A. 2010,
  \bibinfo{title}{DICHOTOMY OF SOLAR CORONAL JETS: STANDARD JETS AND BLOWOUT
  JETS,} The Astrophysical Journal, 720, 757,
  \dodoi{10.1088/0004-637X/720/1/757}

\bibitem[{D. {M{\"u}ller} {et~al.}(2020){M{\"u}ller}, {St. Cyr}, {Zouganelis},
  {Gilbert}, {Marsden}, {Nieves-Chinchilla}, {Antonucci}, {Auch{\`e}re},
  {Berghmans}, {Horbury}, {Howard}, {Krucker}, {Maksimovic}, {Owen}, {Rochus},
  {Rodriguez-Pacheco}, {Romoli}, {Solanki}, {Bruno}, {Carlsson}, {Fludra},
  {Harra}, {Hassler}, {Livi}, {Louarn}, {Peter}, {Sch{\"u}hle}, {Teriaca}, {del
  Toro Iniesta}, {Wimmer-Schweingruber}, {Marsch}, {Velli}, {De Groof},
  {Walsh}, \& {Williams}}]{2020A&A...642A...1M}
{M{\"u}ller}, D., {St. Cyr}, O.~C., {Zouganelis}, I., {et~al.} 2020,
  \bibinfo{title}{{The Solar Orbiter mission. Science overview},} \aap, 642,
  A1, \dodoi{10.1051/0004-6361/202038467}

\bibitem[{Z. {Ning}(2008){Ning}}]{2008ApJ...686..674N}
{Ning}, Z. 2008, \bibinfo{title}{{RHESSI Microflares with Quiet Microwave
  Emission},} \apj, 686, 674, \dodoi{10.1086/590652}

\bibitem[{P. {Pagano} {et~al.}(2021){Pagano}, {Antolin}, \&
  {Petralia}}]{2021A&A...656A.141P}
{Pagano}, P., {Antolin}, P., \& {Petralia}, A. 2021, \bibinfo{title}{{Modelling
  of asymmetric nanojets in coronal loops},} \aap, 656, A141,
  \dodoi{10.1051/0004-6361/202141030}

\bibitem[{E. {Pariat} {et~al.}(2005){Pariat}, {D{\'e}moulin}, \&
  {Berger}}]{2005A&A...439.1191P}
{Pariat}, E., {D{\'e}moulin}, P., \& {Berger}, M.~A. 2005,
  \bibinfo{title}{{Photospheric flux density of magnetic helicity},} \aap, 439,
  1191, \dodoi{10.1051/0004-6361:20052663}

\bibitem[{E.~N. {Parker}(1988){Parker}}]{1988ApJ...330..474P}
{Parker}, E.~N. 1988, \bibinfo{title}{{Nanoflares and the Solar X-Ray Corona},}
  \apj, 330, 474, \dodoi{10.1086/166485}

\bibitem[{R. {Patel} \& V. {Pant}(2022){Patel} \& {Pant}}]{2022ApJ...938..122P}
{Patel}, R., \& {Pant}, V. 2022, \bibinfo{title}{{Hi-C 2.1 Observations of
  Reconnection Nanojets},} \apj, 938, 122, \dodoi{10.3847/1538-4357/ac92e5}

\bibitem[{W.~D. {Pesnell} {et~al.}(2012){Pesnell}, {Thompson}, \&
  {Chamberlin}}]{2012SoPh..275....3P}
{Pesnell}, W.~D., {Thompson}, B.~J., \& {Chamberlin}, P.~C. 2012,
  \bibinfo{title}{{The Solar Dynamics Observatory (SDO)},} \solphys, 275, 3,
  \dodoi{10.1007/s11207-011-9841-3}

\bibitem[{J. {Plowman} \& A. {Caspi}(2020){Plowman} \&
  {Caspi}}]{2020ApJ...905...17P}
{Plowman}, J., \& {Caspi}, A. 2020, \bibinfo{title}{{A Fast, Simple, Robust
  Algorithm for Coronal Temperature Reconstruction},} \apj, 905, 17,
  \dodoi{10.3847/1538-4357/abc260}

\bibitem[{D.~I. {Pontin} {et~al.}(2017){Pontin}, {Janvier}, {Tiwari},
  {Galsgaard}, {Winebarger}, \& {Cirtain}}]{2017ApJ...837..108P}
{Pontin}, D.~I., {Janvier}, M., {Tiwari}, S.~K., {et~al.} 2017,
  \bibinfo{title}{{Observable Signatures of Energy Release in Braided Coronal
  Loops},} \apj, 837, 108, \dodoi{10.3847/1538-4357/aa5ff9}

\bibitem[{J. {Qiu} {et~al.}(2004){Qiu}, {Liu}, {Gary}, {Nita}, \&
  {Wang}}]{2004ApJ...612..530Q}
{Qiu}, J., {Liu}, C., {Gary}, D.~E., {Nita}, G.~M., \& {Wang}, H. 2004,
  \bibinfo{title}{{Hard X-Ray and Microwave Observations of Microflares},}
  \apj, 612, 530, \dodoi{10.1086/422401}

\bibitem[{N.~E. {Raouafi} {et~al.}(2016){Raouafi}, {Patsourakos}, {Pariat},
  {Young}, {Sterling}, {Savcheva}, {Shimojo}, {Moreno-Insertis}, {DeVore},
  {Archontis}, {T{\"o}r{\"o}k}, {Mason}, {Curdt}, {Meyer}, {Dalmasse}, \&
  {Matsui}}]{2016SSRv..201....1R}
{Raouafi}, N.~E., {Patsourakos}, S., {Pariat}, E., {et~al.} 2016,
  \bibinfo{title}{{Solar Coronal Jets: Observations, Theory, and Modeling},}
  \ssr, 201, 1, \dodoi{10.1007/s11214-016-0260-5}

\bibitem[{J. {Reid} {et~al.}(2018){Reid}, {Hood}, {Parnell}, {Browning}, \&
  {Cargill}}]{2018A&A...615A..84R}
{Reid}, J., {Hood}, A.~W., {Parnell}, C.~E., {Browning}, P.~K., \& {Cargill},
  P.~J. 2018, \bibinfo{title}{{Coronal energy release by MHD avalanches:
  continuous driving},} \aap, 615, A84, \dodoi{10.1051/0004-6361/201732399}

\bibitem[{P. {Rochus} {et~al.}(2020){Rochus}, {Auch{\`e}re}, {Berghmans},
  {Harra}, {Schmutz}, {Sch{\"u}hle}, {Addison}, {Appourchaux}, {Aznar
  Cuadrado}, {Baker}, {Barbay}, {Bates}, {BenMoussa}, {Bergmann}, {Beurthe},
  {Borgo}, {Bonte}, {Bouzit}, {Bradley}, {B{\"u}chel}, {Buchlin},
  {B{\"u}chner}, {Cab{\'e}}, {Cadiergues}, {Chaigneau}, {Chares}, {Choque
  Cortez}, {Coker}, {Condamin}, {Coumar}, {Curdt}, {Cutler}, {Davies},
  {Davison}, {Defise}, {Del Zanna}, {Delmotte}, {Delouille}, {Dolla},
  {Dumesnil}, {D{\"u}rig}, {Enge}, {Fran{\c{c}}ois}, {Fourmond}, {Gillis},
  {Giordanengo}, {Gissot}, {Green}, {Guerreiro}, {Guilbaud}, {Gyo},
  {Haberreiter}, {Hafiz}, {Hailey}, {Halain}, {Hansotte}, {Hecquet},
  {Heerlein}, {Hellin}, {Hemsley}, {Hermans}, {Hervier}, {Hochedez},
  {Houbrechts}, {Ihsan}, {Jacques}, {J{\'e}r{\^o}me}, {Jones}, {Kahle},
  {Kennedy}, {Klaproth}, {Kolleck}, {Koller}, {Kotsialos}, {Kraaikamp},
  {Langer}, {Lawrenson}, {Le Clech'}, {Lenaerts}, {Liebecq}, {Linder}, {Long},
  {Mampaey}, {Markiewicz-Innes}, {Marquet}, {Marsch}, {Matthews}, {Mazy},
  {Mazzoli}, {Meining}, {Meltchakov}, {Mercier}, {Meyer}, {Monecke}, {Monfort},
  {Morinaud}, {Moron}, {Mountney}, {M{\"u}ller}, {Nicula}, {Parenti}, {Peter},
  {Pfiffner}, {Philippon}, {Phillips}, {Plesseria}, {Pylyser}, {Rabecki},
  {Ravet-Krill}, {Rebellato}, {Renotte}, {Rodriguez}, {Roose}, {Rosin},
  {Rossi}, {Roth}, {Rouesnel}, {Roulliay}, {Rousseau}, {Ruane}, {Scanlan},
  {Schlatter}, {Seaton}, {Silliman}, {Smit}, {Smith}, {Solanki}, {Spescha},
  {Spencer}, {Stegen}, {Stockman}, {Szwec}, {Tamiatto}, {Tandy}, {Teriaca},
  {Theobald}, {Tychon}, {van Driel-Gesztelyi}, {Verbeeck}, {Vial}, {Werner},
  {West}, {Westwood}, {Wiegelmann}, {Willis}, {Winter}, {Zerr}, {Zhang}, \&
  {Zhukov}}]{2020A&A...642A...8R}
{Rochus}, P., {Auch{\`e}re}, F., {Berghmans}, D., {et~al.} 2020,
  \bibinfo{title}{{The Solar Orbiter EUI instrument: The Extreme Ultraviolet
  Imager},} \aap, 642, A8, \dodoi{10.1051/0004-6361/201936663}

\bibitem[{J. {Saqri} {et~al.}(2022){Saqri}, {Veronig}, {Warmuth}, {Dickson},
  {Battaglia}, {Podladchikova}, {Xiao}, {Battaglia}, {Hurford}, \&
  {Krucker}}]{2022A&A...659A..52S}
{Saqri}, J., {Veronig}, A.~M., {Warmuth}, A., {et~al.} 2022,
  \bibinfo{title}{{Multi-instrument STIX microflare study},} \aap, 659, A52,
  \dodoi{10.1051/0004-6361/202142373}

\bibitem[{P.~W. {Schuck}(2006){Schuck}}]{2006ApJ...646.1358S}
{Schuck}, P.~W. 2006, \bibinfo{title}{{Tracking Magnetic Footpoints with the
  Magnetic Induction Equation},} \apj, 646, 1358, \dodoi{10.1086/505015}

\bibitem[{Y. {Shen}(2021){Shen}}]{2021RSPSA.47700217S}
{Shen}, Y. 2021, \bibinfo{title}{{Observation and modelling of solar jets},}
  Proceedings of the Royal Society of London Series A, 477, 217,
  \dodoi{10.1098/rspa.2020.0217}

\bibitem[{K. {Shibata} {et~al.}(1989){Shibata}, {Tajima}, {Steinolfson}, \&
  {Matsumoto}}]{1989ApJ...345..584S}
{Shibata}, K., {Tajima}, T., {Steinolfson}, R.~S., \& {Matsumoto}, R. 1989,
  \bibinfo{title}{{Two-dimensional magnetohydrodynamic model of emerging
  magnetic flux in the solar atmosphere},} \apj, 345, 584,
  \dodoi{10.1086/167932}

\bibitem[{K. {Shibata} {et~al.}(1992){Shibata}, {Ishido}, {Acton}, {Strong},
  {Hirayama}, {Uchida}, {McAllister}, {Matsumoto}, {Tsuneta}, {Shimizu},
  {Hara}, {Sakurai}, {Ichimoto}, {Nishino}, \& {Ogawara}}]{1992PASJ...44L.173S}
{Shibata}, K., {Ishido}, Y., {Acton}, L.~W., {et~al.} 1992,
  \bibinfo{title}{{Observations of X-Ray Jets with the YOHKOH Soft X-Ray
  Telescope},} \pasj, 44, L173

\bibitem[{T. Shimizu {et~al.}(2002)Shimizu, Shine, Title, Tarbell, \&
  Frank}]{Shimizu_2002}
Shimizu, T., Shine, R.~A., Title, A.~M., Tarbell, T.~D., \& Frank, Z. 2002,
  \bibinfo{title}{Photospheric Magnetic Activities Responsible for Soft X-Ray
  Pointlike Microflares. I. Identifications of Associated
  Photospheric/Chromospheric Activities,} The Astrophysical Journal, 574, 1074,
  \dodoi{10.1086/340998}

\bibitem[{M. {Shimojo} {et~al.}(1996){Shimojo}, {Hashimoto}, {Shibata},
  {Hirayama}, {Hudson}, \& {Acton}}]{1996PASJ...48..123S}
{Shimojo}, M., {Hashimoto}, S., {Shibata}, K., {et~al.} 1996,
  \bibinfo{title}{{Statistical Study of Solar X-Ray Jets Observed with the
  YOHKOH Soft X-Ray Telescope},} \pasj, 48, 123, \dodoi{10.1093/pasj/48.1.123}

\bibitem[{A.~R.~C. {Sukarmadji} {et~al.}(2022){Sukarmadji}, {Antolin}, \&
  {McLaughlin}}]{2022ApJ...934..190S}
{Sukarmadji}, A. R.~C., {Antolin}, P., \& {McLaughlin}, J.~A. 2022,
  \bibinfo{title}{{Observations of Instability-driven Nanojets in Coronal
  Loops},} \apj, 934, 190, \dodoi{10.3847/1538-4357/ac7870}

\bibitem[{J.~Q. {Sun} {et~al.}(2015){Sun}, {Cheng}, {Ding}, {Guo}, {Priest},
  {Parnell}, {Edwards}, {Zhang}, {Chen}, \& {Fang}}]{2015NatCo...6.7598S}
{Sun}, J.~Q., {Cheng}, X., {Ding}, M.~D., {et~al.} 2015,
  \bibinfo{title}{{Extreme ultraviolet imaging of three-dimensional magnetic
  reconnection in a solar eruption},} Nature Communications, 6, 7598,
  \dodoi{10.1038/ncomms8598}

\bibitem[{Z. {Svestka}(1976){Svestka}}]{1976GAM.....8.....S}
{Svestka}, Z. 1976, {Solar flares}, Vol.~8

\bibitem[{Y.~H. {Tang} {et~al.}(2000){Tang}, {Li}, {Fang}, {Aulanier},
  {Schmieder}, {Demoulin}, \& {Sakurai}}]{2000ApJ...534..482T}
{Tang}, Y.~H., {Li}, Y.~N., {Fang}, C., {et~al.} 2000,
  \bibinfo{title}{{H{\ensuremath{\alpha}} and Soft X-Ray Brightening Events
  Caused by Emerging Flux},} \apj, 534, 482, \dodoi{10.1086/308715}

\bibitem[{N.~M. {Viall} \& J.~A. {Klimchuk}(2011){Viall} \&
  {Klimchuk}}]{2011ApJ...738...24V}
{Viall}, N.~M., \& {Klimchuk}, J.~A. 2011, \bibinfo{title}{{Patterns of
  Nanoflare Storm Heating Exhibited by an Active Region Observed with Solar
  Dynamics Observatory/Atmospheric Imaging Assembly},} \apj, 738, 24,
  \dodoi{10.1088/0004-637X/738/1/24}

\bibitem[{N.~M. {Viall} \& J.~A. {Klimchuk}(2012){Viall} \&
  {Klimchuk}}]{2012ApJ...753...35V}
{Viall}, N.~M., \& {Klimchuk}, J.~A. 2012, \bibinfo{title}{{Evidence for
  Widespread Cooling in an Active Region Observed with the SDO Atmospheric
  Imaging Assembly},} \apj, 753, 35, \dodoi{10.1088/0004-637X/753/1/35}

\bibitem[{H. {Wang} {et~al.}(1999){Wang}, {Chae}, {Qiu}, {Lee}, \&
  {Goode}}]{1999SoPh..188..365W}
{Wang}, H., {Chae}, J., {Qiu}, J., {Lee}, C.-Y., \& {Goode}, P.~R. 1999,
  \bibinfo{title}{{Studies of Microflares and C5.2 flare of 27 September
  1998},} \solphys, 188, 365, \dodoi{10.1023/A:1005288217459}

\bibitem[{W. {Wang} {et~al.}(2016){Wang}, {Wang}, {Krucker}, \&
  {Hannah}}]{2016SoPh..291.1357W}
{Wang}, W., {Wang}, L., {Krucker}, S., \& {Hannah}, I. 2016,
  \bibinfo{title}{{Simulation of Quiet-Sun Hard X-Rays Related to Solar Wind
  Superhalo Electrons},} \solphys, 291, 1357, \dodoi{10.1007/s11207-016-0916-z}

\bibitem[{T. {Yokoyama} \& K. {Shibata}(1996){Yokoyama} \&
  {Shibata}}]{1996PASJ...48..353Y}
{Yokoyama}, T., \& {Shibata}, K. 1996, \bibinfo{title}{{Numerical Simulation of
  Solar Coronal X-Ray Jets Based on the Magnetic Reconnection Model},} \pasj,
  48, 353, \dodoi{10.1093/pasj/48.2.353}

\bibitem[{J. {Zhang} {et~al.}(2012){Zhang}, {Cheng}, \&
  {Ding}}]{2012NatCo...3..747Z}
{Zhang}, J., {Cheng}, X., \& {Ding}, M.-D. 2012, \bibinfo{title}{{Observation
  of an evolving magnetic flux rope before and during a solar eruption},}
  Nature Communications, 3, 747, \dodoi{10.1038/ncomms1753}

\end{thebibliography}

\bibliographystyle{aasjournal}

\begin{deluxetable*}{cccccccccc}
\tablecaption{Peak temperature and heat release \label{tab:messier}}

\tablewidth{0pt}
\tablehead{
\colhead{Flare} & \colhead{} & \colhead{Peak Time} & \colhead{} & \colhead{$T_{P1}$} & \colhead{} & \colhead{$T_{P2}$} & \colhead{} & \colhead{} & \colhead{$E_{th}$} \\
\colhead{} & \colhead{} & \colhead{(UT)} & \colhead{} & \colhead{(MK)} & \colhead{} & \colhead{(MK)} & \colhead{} & \colhead{} & \colhead{(erg)}
}
\startdata
Flare 1 & & 00:44:10 & & 1.8 & & 1.6 & & & $7.5 \times 10^{27}$  \\
Flare 2 & & 00:47:58 & & 1.8 & & 2.0 & & & $7.6 \times 10^{28}$  \\
Flare 3 & & 00:53:10 & & 1.8 & & 2.0 & & & $1.96 \times 10^{29}$  \\
Flare 4 & & 00:59:10 & & 1.8 & & 2.5 & & & $1.25 \times 10^{29}$  \\
\enddata
\tablecomments{\(T_{P1}\) is the background peak temperature and \(T_{P2}\) is the event peak temperature.} 
\end{deluxetable*}
\end{document}